\documentclass[%
 reprint,
 amsmath,amssymb,
 aps,
]{revtex4-2}

\usepackage{graphicx}
\usepackage{dcolumn}
\usepackage{bm}
\DeclareMathOperator{\Tr}{Tr}

\begin{document}


\title{Active nematic-isotropic interfaces on flat surfaces: effects of anchoring, ordering field and activity}
\author{Rodrigo C. V. Coelho} 
\email{rcvcoelho@fc.ul.pt}%
 \author{José A. Moreira}%
\author{Duarte M. C. Pedro}
\author{Margarida M. Telo da Gama} 
\affiliation{Centro de Física Teórica e Computacional, Faculdade de Ciências, Universidade de Lisboa, 1749-016 Lisboa, Portugal.}
\affiliation{Departamento de Física, Faculdade de Ciências, Universidade de Lisboa, P-1749-016 Lisboa, Portugal.}



\begin{abstract}
A surface in contact with the isotropic phase of a passive liquid crystal can induce nematic order over distances that range from microscopic to macroscopic when the nematic-isotropic interface undergoes an orientational-wetting transition. If the nematic is active, what happens to the interface? Does it propagate and, if it does, is its structure different from the passive one? In this paper, we address these questions. We investigate how the active nematic-isotropic interface is affected by the anchoring strength of the surface, the bulk ordering field and the activity. We find that while passive interfaces are one-dimensional the active ones exhibit two dynamical regimes: a passive-like regime and a propagating regime where the interfaces propagate until the entire domain is active nematic. Active interfaces break the translational symmetry within the interfacial plane above a threshold activity, where the active nematic fluctuations, which are ultimately responsible for the emergence of an active turbulent nematic phase, drive non-steady dynamical interfacial regimes.
\end{abstract}

\maketitle


\section{Introduction}

The behaviour of passive nematic films on flat and structured surfaces has been studied extensively, not least because the anchoring and anchoring transitions in these systems play a key role in various applications, notably in display technologies~\cite{Chen2011,doi:10.1080/00268976.2010.542780, p1995physics}. The behaviour of active nematic films on surfaces has been investigated more recently but remains largely unexplored. One of the early works addressed the structure of an extensile active nematic at a planar surface in two dimensions (2D) and revealed that as the thickness of the nematic film grows the nematic order is destroyed by the onset of a bending instability at the free interface~\cite{C7SM00325K,PhysRevLett.113.248303}. More recent studies considered the behaviour of confined active nematics~\cite{doi:10.1126/science.aal1979, SagusMestre2022}. For small active droplets on planar surfaces, a range of dynamical behaviours was reported~\cite{PhysRevResearch.5.033165} and for active nematic-isotropic interfaces in quasi-one dimensional (1D) channels an interfacial instability was observed before the onset of the bulk active turbulent regime~\cite{C9SM00859D}. Other works on active nematic droplets considered their morphology~\cite{PhysRevX.11.021001,Alert_2022} and dynamics~\cite{Tiribocchi2023, PhysRevLett.112.147802}. 
Recently, experiments and theory have been used to address the question of how the activity
controls the interfaces that separate an active from a passive
fluid~\cite{doi:10.1126/science.abo5423}. The authors reported that when in contact
with a solid surface, the active-passive interface exhibits a
nonequilibrium wetting transition.

However, a systematic study of the effects of confinement, surface anchoring, temperature (ordering field) and activity on the active nematic-isotropic interface is still missing. While the confinement and surface anchoring promote nematic order, the activity promotes a globally disordered active turbulent state. The ordering field may promote either global order or disorder by stabilising the nematic or the isotropic phases in the bulk.   
In what follows we address these questions for 1D and 2D (extensile) active nematic films on surfaces. The surface anchoring promotes nematic ordering while the ordering field determines the thickness of the ordered film in contact with the surface as in passive systems. The activity will ultimately destabilize a 2D film by promoting a dynamical state, where the interface propagates and the uniform nematic is replaced by a non-uniform active nematic state, which evolves in time and becomes pre-turbulent or turbulent at long times, depending on the activity and the system size. 
We start by revisiting briefly the passive case, where the phenomenology is well established~\cite{sluckin1986orientational}, using the Beris-Edwards equation and the hydrodynamic theory of active nematics based on the Landau-de Gennes theory. This allows us to establish the relevant parameters and provides a stringent test of the numerical accuracy of the method, which is known to be plagued by spurious currents under non-uniform conditions that are inevitable near the surface and at the edge of the free nematic film~\cite{refId0Coelho}. We proceed to study 1D active interfaces, under strong anchoring conditions and reveal that even under strict confinement the activity leads to new dynamical behaviours. These dynamical states are further investigated by considering 2D systems, where the steady states are found to depend strongly on the activity, the system size, the surface and interfacial anchoring and the initial conditions. 

The paper is arranged as follows. Section~\ref{method-sec} describes the model used to study the interfaces and the numerical method. Section~\ref{results-sec} discusses the results for the interfaces with different activities, anchorings and temperatures. Both the 1D and 2D cases are discussed. Section~\ref{conclusion-sec} summarizes and concludes.

\section{Theory and method}
\label{method-sec}

In this section, we summarize the Beris-Edwards model which was used to study the dynamics of the nematic-isotropic interface. We obtain the equation for the interfacial profile and describe the numerical method.

\subsection{Beris-Edwards model}

A uniaxial nematic is characterized by the orientational order parameter $Q_{\alpha \beta} = S \left ( n_{\alpha} n_{\beta} - \frac{1}{3}\delta_{\alpha \beta} \right )$ where $S$ is the degree of orientational order along the director $\boldsymbol{n}$~\cite{doi2013soft}. In the isotropic phase, $S=0$ and, in the nematic phase $S$ is finite, and its value depends on the temperature.

In order to study the system close to the coexistence between the nematic and isotropic phases, the Landau-de Gennes theory for liquid crystals is employed. The free energy density is~\cite{doi:10.1080/15421407108082773,p1995physics, marenduzzo2007steady} 
\begin{align}
    f_{NI} &= \frac{A_{0}}{2}\left(1-\frac{\gamma}{3}\right) Q_{\alpha \beta}^{2} - \frac{A_{0} \gamma}{3} Q_{\alpha \beta} Q_{\beta \gamma} Q_{\gamma \alpha} \nonumber \\ & + \frac{A_{0} \gamma}{4}\left(Q_{\alpha \beta}^{2}\right)^{2}.
\end{align}
The parameter $\gamma$ sets the bulk ordering field and controls the phase transition; it can be understood as an effective inverse temperature (for thermotropic liquid crystals) or an effective concentration (for lyotropic liquid crystals). The coexistence between the bulk nematic and isotropic phases occurs at $\gamma_{NI} = 2.7$ and $S=S_N = 1/3$ as obtained by minimizing $f_{NI}$. For $\gamma > 2.7$ the bulk equilibrium phase is nematic while for $\gamma < 2.7$ the equilibrium phase is isotropic. 

In the bulk nematic state, a configuration with a uniform director field is favourable. However, such a configuration may not be possible due to constraints on the system, such as the interaction with a surface or with an electric field. The spatial variations of the orientational order are penalized with a cost described by the elastic free energy density:
\begin{align}
   f_{E}=\frac{1}{2} K\left(\partial_{\gamma} Q_{\alpha \beta}\right)^{2},
 \label{elastic}
\end{align}
where we assumed the usual one-elastic constant ($K$) approximation.

To take into account surface effects, we introduce a surface anchoring term:
\begin{align} \label{anchoring}
    f_{W}=\frac{1}{2} W\left(Q_{\alpha \beta}-Q_{\alpha \beta}^{0}\right)^{2}
\end{align}
which penalises deviations from the surface-preferred order parameter $Q_{\alpha \beta}^{0}$ with strength $W$. This term is applied only at the surface (fluid in contact with the solid). The total free energy is thus the sum of these three contributions: $\mathcal{F}=\int d^3 x (f_{NI}+f_E) +\int d^2 x f_W$.

The time evolution of the order parameter $Q_{\alpha\beta}$ and of the velocity field is described by the Beris-Edwards equation for the hydrodynamics of liquid crystals~\cite{beris1994thermodynamics}. The Beris-Edwards equation is given by: 
\begin{align}
  \partial _t Q_{\alpha \beta} + u _\gamma \partial _\gamma Q_{\alpha \beta} - S_{\alpha \beta} = \Gamma H_{\alpha\beta} , 
  \label{beris-edwards}
\end{align}
where the molecular field $H_{\alpha\beta}$ is:
\begin{align}
 H_{\alpha\beta} = -\frac{\delta \mathcal{F}}{\delta Q_{\alpha\beta}} + \frac{\delta_{\alpha\beta}}{3} \Tr \left( \frac{\delta \mathcal{F}}{\delta Q_{\gamma \epsilon}} \right),
\end{align}
and the co-rotational term $S_{\alpha \beta}$ reads:
\begin{align}
 S_{\alpha \beta} =& ( \xi D_{\alpha \gamma} + W_{\alpha \gamma})\left(Q_{\beta\gamma} + \frac{\delta_{\beta\gamma}}{3} \right) \\
 & +\left( Q_{\alpha\gamma}+\frac{\delta_{\alpha\gamma}}{3} \right)(\xi D_{\gamma\beta}-W_{\gamma\beta}) \nonumber \\
 & -2\xi\left( Q_{\alpha\beta}+\frac{\delta_{\alpha\beta}}{3}  \right)(Q_{\gamma\epsilon} \partial _\gamma u_\epsilon).
\end{align}
The vorticity and the shear rate are, respectively, $W_{\alpha\beta} = (\partial _\beta u_\alpha - \partial _\alpha u_\beta )/2$ and $D_{\alpha\beta} = (\partial _\beta u_\alpha + \partial _\alpha u_\beta )/2$.
The parameter $\Gamma$ is the rotational diffusion constant and it sets the time scale of the dynamics of the director field and $\xi$ is the aligning parameter, which depends on the particle shape: it is positive for rod-like particles and negative for disk-like ones.

The Navier-Stokes and continuity equations, which govern the dynamics of the velocity field, read:
\begin{align}
&\rho \partial _t u_\alpha + \rho u_\beta \partial_\beta u_{\alpha} = \partial _\beta \Pi_{\alpha\beta}  + \eta \partial_\beta\left( \partial_\alpha u_\beta + \partial_\beta u_\alpha \right),\nonumber \\
&\partial_\alpha  u_\alpha =0.
\label{navier-stokes}
\end{align}
The stress tensor $\Pi_{\alpha\beta}$ is given by the sum of the passive stress tensor, given by
\begin{align} 
 \Pi^{\text{passive}}_{\alpha\beta} =& -P_0 \delta_{\alpha\beta} + 2\xi \left( Q_{\alpha\beta} +\frac{\delta_{\alpha\beta}}{3} \right)Q_{\gamma\epsilon}H_{\gamma\epsilon} \nonumber \\ &- \xi H_{\alpha\gamma} \left( Q_{\gamma\beta}+\frac{\delta_{\gamma\beta}}{3} \right) - \xi \left( Q_{\alpha\gamma} +\frac{\delta_{\alpha\gamma}}{3} \right) H_{\gamma \beta} \nonumber \\ &- \partial _\alpha Q_{\gamma\nu} \,\frac{\delta \mathcal{F}}{\delta (\partial_\beta Q_{\gamma\nu})} + Q_{\alpha\gamma}H_{\gamma\beta} - H_{\alpha\gamma}Q_{\gamma\beta},
 \label{passive-pressure-eq}
\end{align}
and the active stress tensor is proportional to the nematic tensor order parameter~\cite{PhysRevLett.89.058101,doostmohammadi2018active}:
\begin{align}
 \Pi^{\text{active}}_{\alpha\beta} = -\zeta Q_{\alpha\beta}.
 \label{active-pressure-eq}
\end{align}
In Eq.~\eqref{navier-stokes}, $\eta$ is the absolute viscosity, and in Eq.~\eqref{passive-pressure-eq}, $P_0$ is the hydrostatic pressure. The parameter $\zeta$ controls the activity level, being positive for extensile systems and negative for contractile ones. We will consider extensile systems. A useful parameter to interpret the results is the active length $\ell_A = \sqrt{K/\zeta}$, which governs the size of the vortices and the distance between defects. For instance, as reported in Ref.~\cite{C9SM02306B} (for a similar model, with friction) the vortex size is $\ell_v \approx 10 \ell_A$.

\subsection{Nematic-isotropic interface at equilibrium}

Here we will obtain a semi-analytical expression for the scalar order parameter $S$ across the nematic-isotropic interface at equilibrium. Thus, we will consider a passive system, with the velocity field being zero everywhere, in the steady state.

Let us take the nematic system in contact with a surface on the $xy$ plane that induces homeotropic (normal) anchoring. We want to find the equilibrium state $Q_{\alpha \beta}^{eq}$ that minimizes the free energy discussed in the previous section. We assume infinite anchoring, $W \rightarrow \infty$. The free energy is minimized by solving:
\begin{align}
    \frac{\delta \mathcal{F}^{\text {bulk }}}{\delta Q_{\alpha \beta}}=0.
\end{align}
The ordering field $\gamma$ is fixed at a value close to coexistence in the isotropic phase ($\gamma \lesssim 2.7$). The surface induces nematic order while $\gamma$ favours the isotropic phase. Thus, an interface sets in between the two states at a certain distance, and height, from the surface. The equilibrium interfacial height is determined by the distance from the nematic-isotropic coexistence and the anchoring strength.   

We use Beltrami's identity to solve for $Q_{\alpha \beta}$:
\begin{align}
    f-\nabla Q_{\alpha \beta} \cdot \frac{\partial f}{\partial \nabla Q_{\alpha \beta}}=c   , 
\end{align}
where $f=f_{N I}+f_{E}$ and $c$ is a constant that needs to be determined. Only the elastic contribution, $f_{E}$, depends on $\nabla Q_{\alpha \beta}$, so we have:
\begin{align}
    f_{NI}-\frac{3}{2} K\left(\nabla Q_{\alpha \beta}\right)^{2}=c
\end{align}
After a few simple steps one finds $Q_{\alpha \beta} Q_{\alpha \beta}=Q_{\alpha \beta}^{2}=\frac{2}{3} S^{2}$,  $Q_{\alpha \beta} Q_{\beta \gamma} Q_{\gamma \alpha}=\frac{2}{9} S^{3}$, and $Q_{\alpha\beta}^4=\frac{4}{9}S^4$, which, when included in $f_{N I}$, give:
\begin{align}
    f_{N I}=\frac{A_{0}}{3}\left(1-\frac{\gamma}{3}\right) S^{2}-\frac{2 A_{0} \gamma}{27} S^{3}+\frac{A_{0} \gamma}{9} S^{4}.
\end{align}
Due to the symmetry of the problem $Q_{\alpha \beta}(\textbf{r})=Q_{\alpha \beta}(z)$, which implies $\nabla Q_{\alpha \beta}=\frac{d Q_{\alpha \beta}}{d z} \hat{\textbf{e}}_{z}$.

For positive $W$ the surface anchoring induces homeotropic nematic order with $\textbf{n}=\hat{\textbf{e}}_z$. Since the elastic free energy penalizes distortions in the director field, the state of minimum free energy has $\textbf{n}$ constant and equal to $\hat{\textbf{e}}_z$, while $S$ can vary in space. Therefore:
\begin{align}
    \frac{d Q_{\alpha \beta}}{d y}=\frac{d S}{d y}\left(n_{\alpha} n_{\beta}-\frac{1}{3} \delta_{\alpha \beta}\right),
\end{align}
and the differential equation for $S$ becomes:
\begin{align}
    f_{N I}-K\left(\frac{d S}{d z}\right)^{2}=c    .
\end{align}

Far from the surface, the liquid crystal is in the isotropic state, i.e.:
\begin{align}
    \left.\frac{d S}{d z}\right|_{z=\infty}=\left.S\right|_{z=\infty} =0.
    \label{bc-eq-inf}
\end{align}

At infinity:
\begin{align}
    f_{N I}(S=0)=0=c    ,
\end{align}
and the equation for $S$ reads:
\begin{align}
    \frac{A_{0}}{3}\left(1-\frac{\gamma}{3}\right) S^{2}(z)-\frac{2 A_{0} \gamma}{27} S^{3}(z)+\frac{A_{0} \gamma}{9} S^{4}(z)=K\left(S^{\prime}(z)\right)^{2}
    \label{eq-s-an}
\end{align}
This is a first-order ordinary differential equation with boundary conditions given by $S(0)=S_{0}$ and Eq.~\eqref{bc-eq-inf}. We take $S_0$ the value of $S$ at the nematic-isotropic transition $S_N=1/3$.
Equation~\eqref{eq-s-an} is solved numerically to obtain the profile of $S$ at equilibrium, which will be compared with the results from the Beris-Edwards equation in the steady state.

\subsection{Numerical method}

We use a hybrid method to solve the hydrodynamic equations of nematics, as done in previous works~\cite{marenduzzo2007steady, C9SM00859D}. The Beris-Edwards equation, Eq.\eqref{beris-edwards}, is solved using finite differences. The derivatives are calculated using central differences of second-order accuracy in space while the time derivative is calculated using a predictor-corrector algorithm. The Navier-Stokes equation, Eq.~\eqref{navier-stokes}, is recovered in the macroscopic limit using the lattice Boltzmann method~\cite{Kruger2017,succi2018lattice}. We use a basic algorithm for fluid flow with the extra terms in Eq.~\eqref{passive-pressure-eq} and \eqref{active-pressure-eq} being implemented as force terms. All the gradients in the force term are calculated using finite differences as before. The quantities and parameters in this paper are in lattice units: the lattice spacing $\Delta x$, the time step $\Delta t$ and the reference density $\rho_r$ are equal to one.

The parameters used in the simulations are as follows: density $\rho=10$, absolute viscosity $\eta=1.67$ (relaxation time $\tau=1$), aligning parameter $\xi=0.8$ (flow aligning regime), elastic constant $K=0.04$, free energy constant $A_0=0.1$, and rotational diffusion constant $\Gamma=0.34$. The values of the activity $\zeta$, surface anchoring $W$ and temperature $\gamma$ are indicated in the corresponding sections and figures.

\section{Results}
\label{results-sec}

In this section, we discuss the results from the simulations for passive and active interfaces.

\subsection{Passive interface}

We start by analyzing a passive nematic-isotropic interface on a flat surface. The domain is one-dimensional ($L_x=1$) with height $L_z=64$. Periodic boundary conditions are applied in the $x$-direction, which means that there are no gradients in this direction. We have tested wider systems but there were no differences when compared to the 1D simulations of passive nematics. The active case will be discussed in the next section, including the results of 2D simulations. At the top and bottom, there are solid surfaces with imposed nematic anchoring at the bottom (homeotropic, with $S=S_0$) and isotropic ($S=0$) at the top. No slip conditions are imposed on both surfaces.

\begin{figure}[h]
    \includegraphics[width=0.45\textwidth]{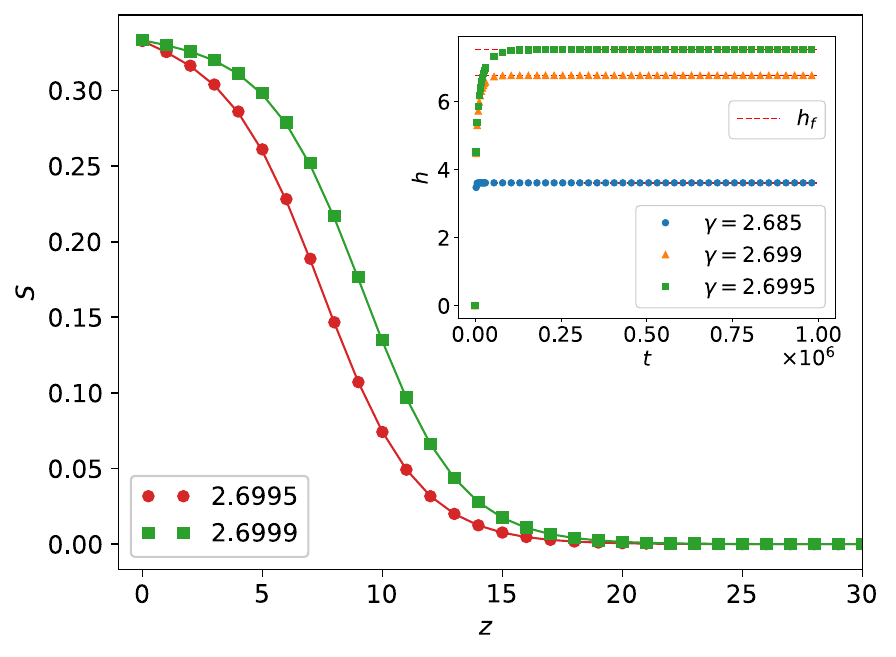}
    \caption{Comparison between the semi-analytic order parameter $S$ profile (solid lines) and the steady state profile obtained from the dynamical simulations for two values of $\gamma$. The inset depicts the time evolution of the film thickness $h$ for three different values of $\gamma$. The dashed lines in the inset indicate the final equilibrium interfacial height or film thickness.}
    \label{final_height-fig}
\end{figure}

The liquid crystal starts at rest in the isotropic phase and quickly an interface is formed between the isotropic phase at the top and the nematic state induced by the surface at the bottom. First, as a validation of the numerics, we compare the interfacial profile of $S$ given by Eq.~\eqref{eq-s-an} with the results from the simulations in the steady state. Figure~\ref{final_height-fig} compares the results for two values of $\gamma$, showing that the simulation results are indistinguishable from those of the semi-analytic profile. The inset of Fig.~\ref{final_height-fig} indicates that the interface converges rapidly to its equilibrium height and stays there in the steady state. 

\begin{figure}[h]
\includegraphics[width=0.45\textwidth]{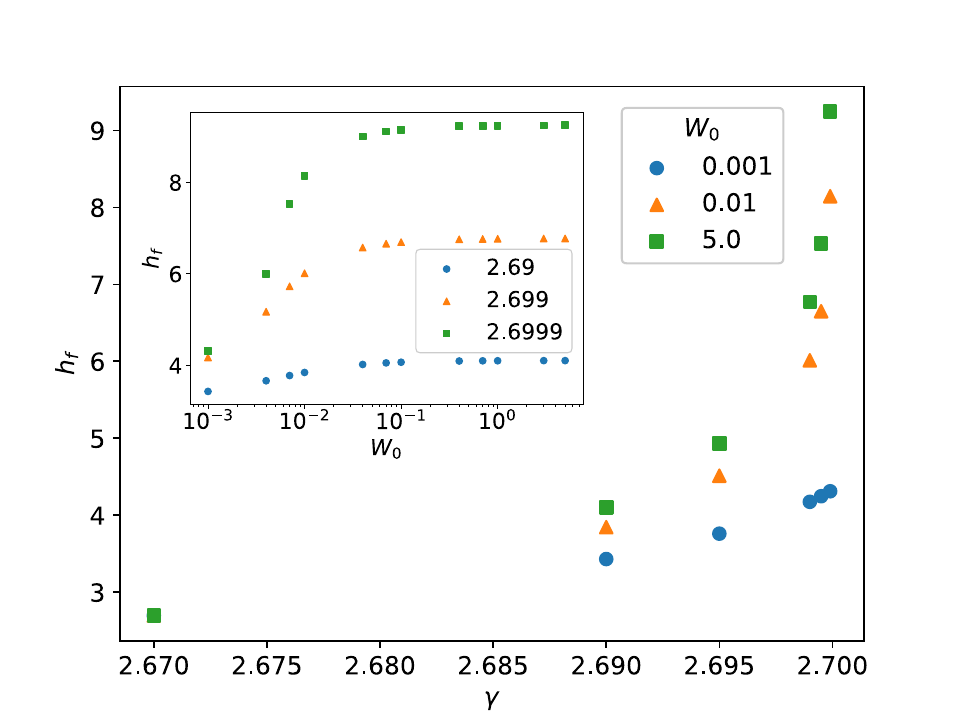}
\caption{ Final equilibrium height $h_f$ of the interface as a function of temperature, $\gamma$, for different values of the anchoring strength, $W$. Except for the lowest anchoring $W$ the interfacial height diverges (logarithmically) at the transition temperature, for an infinite system. The inset depicts the final equilibrium height of the interface as a function of, $W$, at different values of the ordering field, $\gamma$.}
\label{hf_w0}
\end{figure}

Next, we investigate the effect of the surface anchoring strength $W$ and of temperature $\gamma$. Figure~\ref{hf_w0} indicates the final equilibrium height as a function of $\gamma$ for different values of $W$. For strong anchoring, the closer the temperature from coexistence $\gamma_{NI}=2.7$, the higher the interface (the thicker the nematic film) in line with the expectations of a logarithmic divergence of the nematic film thickness for infinite systems (orientational wetting)~\cite{FNBraun_1996,PhysRevLett.37.1059, PhysRevA.26.1610}. This means that the surface can induce nematic order deep into the system as the temperature approaches that of the nematic-isotropic transition $\gamma\geq\gamma_{NI}$, where the nematic film becomes macroscopic and the interface unbinds from the surface (orientational wetting). One finds that at fixed temperature the interfacial height increases with $W$. However, as illustrated in the inset, the height saturates at a certain value of $W\sim 0.1$ at any temperature, above the bulk transition. The orientational wetting transition for infinite anchoring will be observed also at finite anchorings above a certain anchoring strength, as the free energy cost of creating the nematic-isotropic interface will have to be offset by the surface energy that favours nematic ordering. These are well-known results and analytical results for the anchoring threshold may be found in~\cite{FNBraun_1996,PhysRevLett.37.1059, PhysRevA.26.1610}.
From now on we will consider infinite anchoring only (in practice well described by $W=5$). 

\subsection{Active interface}

We proceed to analyze the effect of activity at different values of the ordering field, $\gamma$, under infinite anchoring conditions ($W=5$). As will be discussed, the system is essentially 1D in many of the tested cases. The interface becomes 2D when the translational symmetry is broken spontaneously or by the initial conditions (among other necessary conditions such as high activity). 

\subsubsection{One-dimensional interface}

\begin{figure}[h]
\includegraphics[width=0.45\textwidth]{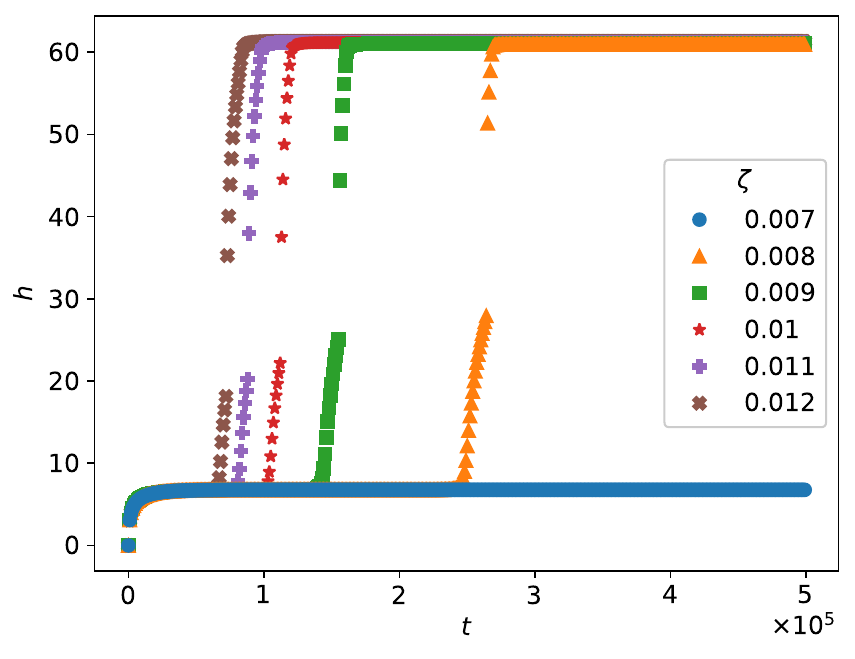}
\caption{ Height of the interface as a function of time for different values of the activity $\zeta$ at $\gamma=2.699$.}
\label{h_t_with_activity}
\end{figure}

We start, as in the previous section with a 1D system, and study the evolution of the interfacial height with time. As shown in Fig.~\ref{h_t_with_activity} the active interfaces initially stabilize at the same height as the passive ones under the same bulk conditions ($h_f \approx 6.8$ at $\gamma=2.699$). Then, after some time that decreases with increasing activity, the interface propagates rapidly until it reaches the top of the domain and the entire system becomes nematic. Thus, we observe two regimes: the active interface stays pinned (finite nematic film) as the passive one, at short times and it propagates reaching the top boundary at long times. In the latter case, the steady state is a highly confined straight channel filled with active nematic, a well-studied system~\cite{marenduzzo2007steady,Hardoin2019, THAMPI2022101613}. In some rare cases, the interface stops in the middle of the 1D domain ($h_f \approx L_z/2$), which was confirmed for other system sizes. This is due to the symmetry of the velocity field in the two halves of the domain, which, in those cases, flows in opposite directions with zero velocity in the middle of the active nematic domain. This is, however, an unstable steady state with nearby configurations flowing away from it to the stable steady state that corresponds to the channel filled with active nematic. 
\begin{figure}[h]
\includegraphics[width=0.45\textwidth]{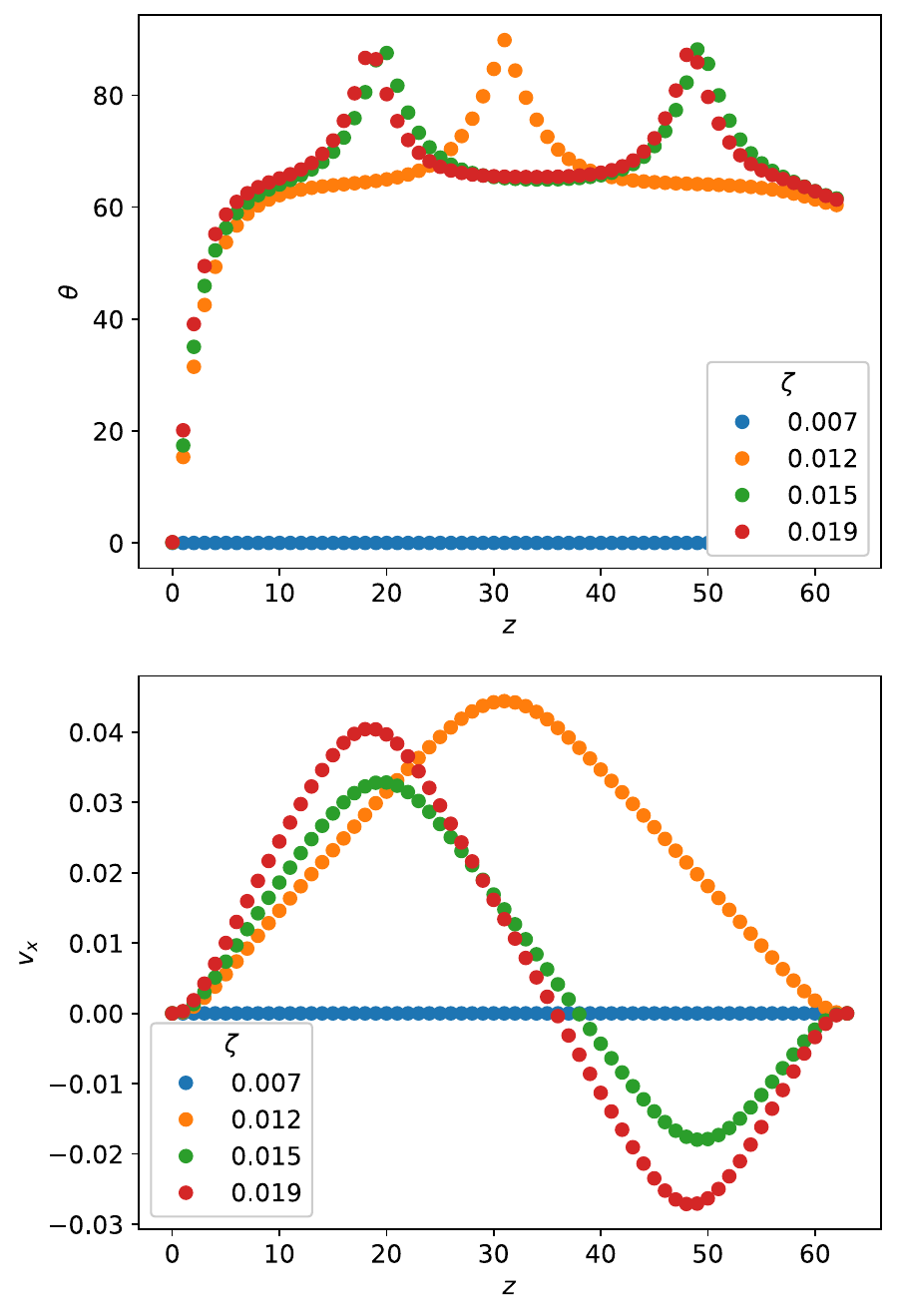}
\caption{(top) Angle in degrees between the director and the vertical along $z$. (Bottom) Velocity field along $z$. Four different activities at $\gamma=2.699$.}
\label{v_n_with_activity}
\end{figure}
Figure~\ref{v_n_with_activity} depicts the angle of the director field with the $z$-axis and the flow fields in the steady state for four different activities at $\gamma=2.699$. At this temperature, the interface propagates above a threshold activity $\zeta^*=0.008$, corresponding to an active length $\ell_A=2.2$ or vortex size of $\sim 25.5$~\cite{C9SM02306B} (here the vortex size can be larger as we do not consider the effect of friction). At low activities, up to $\zeta=0.007$, the directors are aligned with $z$ (following the surface homeotropic anchoring) and the velocity is nearly zero everywhere. This is the thin film or static interface regime.  Above this threshold, the interface propagates and the domain is filled with active nematic. This is the thick film or propagating interface regime. In a finite system the active nematic fills the domain, and at low activities, $\zeta=0.012$ in the figure, there is spontaneous unidirectional flow as observed in active nematics confined in channels. The directors are horizontal in the centre of the domain, where the velocity gradients vanish.  At higher activities, each half of the domain flows in a different direction and the directors are horizontal where the velocity gradients vanish. This two-way flow in 1D, which is the constrained 1D version of 2D vortices, was reported in previous works~\cite{marenduzzo2007steady} although with different anchoring at the walls. In a more realistic 2D description, with initial director fluctuations, the nematic becomes turbulent or pre-turbulent as discussed in the next section~\cite{doi:10.1073/pnas.1202032109, doi:10.1146/annurev-conmatphys-082321-035957, PhysRevFluids.8.124401}. 

\begin{figure}[h]
\includegraphics[width=0.4\textwidth]{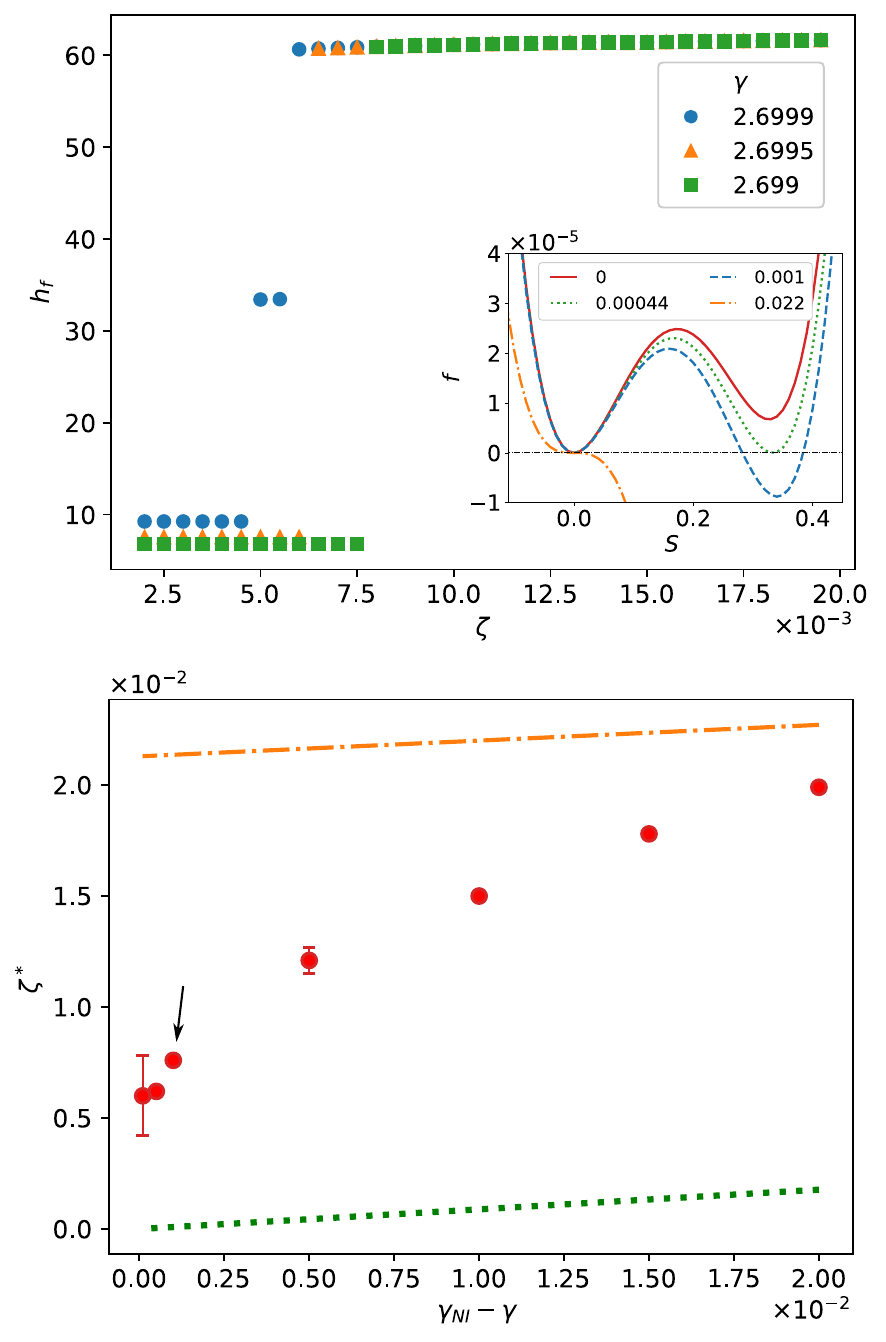}
\caption{(Top) Final height $h_f$ of the interface as a function of the activity $\zeta$ at different values of the ordering field $\gamma$. The inset illustrates the effective free energy density ($f_{NI}+f_A$) at four different activities and $\gamma = 2.695$. The green curve corresponds to the effective free energy for the threshold activity $\zeta_{co}^*$ at this value of $\gamma$. (Bottom) The circles are the threshold activity $\zeta^*$ for interfacial propagation in simulations with isotropic initial conditions,
as a function of the ordering field, $\gamma$. The error bars correspond to the transition interval (for some points the error bars are smaller than the symbol). The green curve (dotted) is threshold activity $\zeta_{co}^*$ where the nematic and isotropic states have the same effective free energy while the orange curve (dash-dotted) is the threshold activity $\zeta_i^*$ where the isotropic state becomes unstable (Eq.~\eqref{zeta2-eq}). The arrow indicates the system at $\gamma=2.699$ analysed in Fig.~\ref{h_t_with_activity}. }
\label{hf_zeta}
\end{figure}
\begin{figure}[h]
\includegraphics[width=0.45\textwidth]{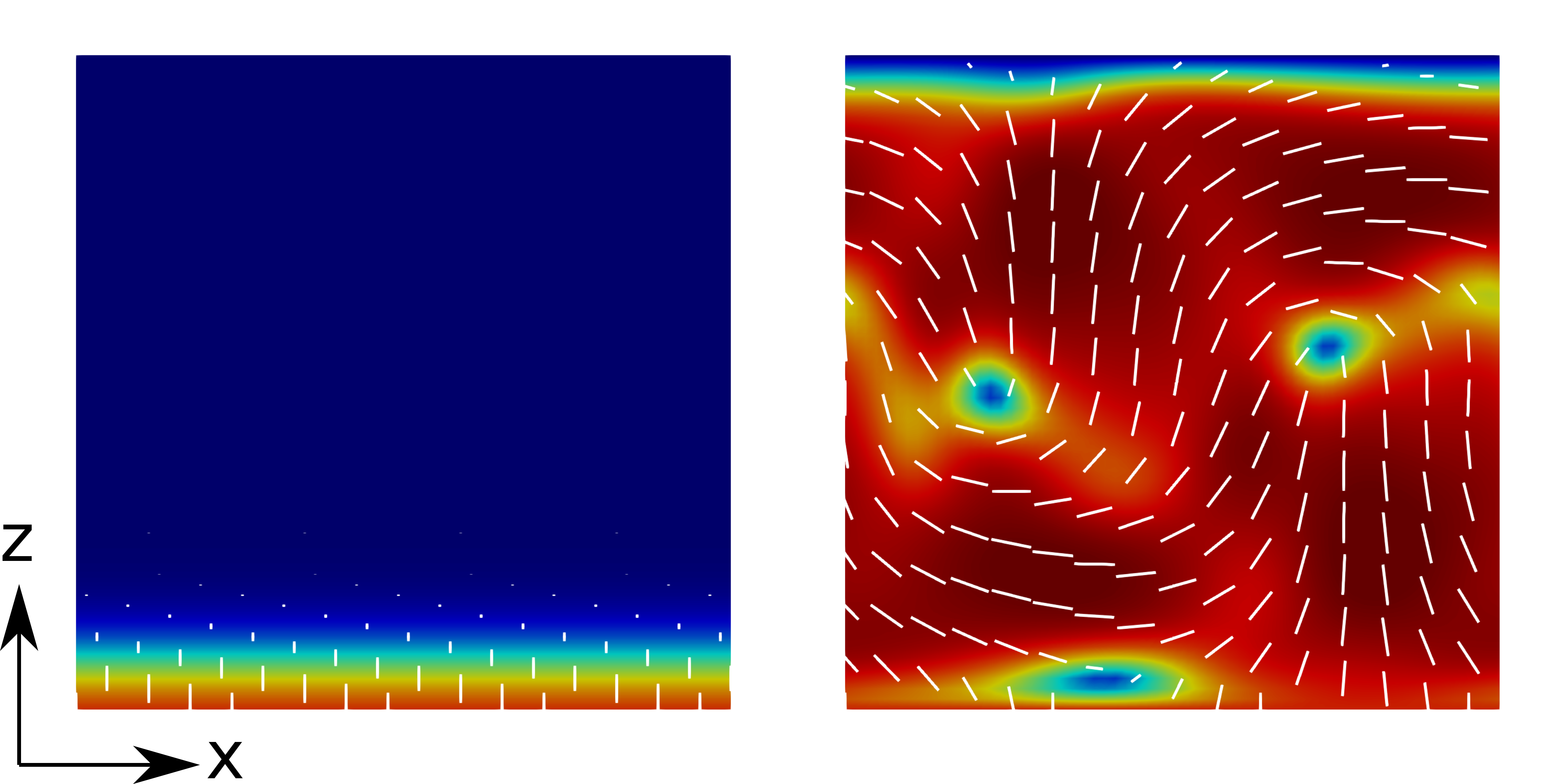}
\caption{Effect of the initial conditions. We set $\zeta=0.009$ and $\gamma=2.695$ in both simulations. On the left, the initial condition is an isotropic fluid everywhere while, on the right, half of the domain is isotropic and the other half is nematic with directors pointing vertically with fluctuations in the director field. Blue and red represent the isotropic and nematic phases respectively while the white lines are the director field.}
\label{2d-int}
\end{figure}

In Fig.~\ref{hf_zeta} (top), the final interfacial height or film thickness (after $5\times 10^6$ iterations) is plotted as a function of the activity, $\zeta$, at three different ordering fields, $\gamma$, in the isotropic phase, close to the bulk nematic-isotropic transition (from 3.7 to 0.37\%). There is a threshold activity, $\zeta^*$, for each value of $\gamma$, required for interfacial propagation with initial isotropic conditions. Figure~\ref{hf_zeta} (bottom) depicts this threshold as a function of $\gamma_{NI}-\gamma$. It reveals that the activity required for interfacial propagation increases as $\gamma$ moves away from the transition temperature. This is to be expected as the activity of extensile systems favours the nematic state and it may be interpreted as generating an effective free energy in addition to the free energy of the passive bulk phases~\cite{Thampi_2015, C8SM02103A}. 

The effective free energy generated by the activity is~\cite{Thampi_2015}: 
\begin{align}
    f_A = -\frac{\xi\zeta}{6 \eta \Gamma} Q_{\alpha \beta}^{2} + \frac{\xi\zeta}{3 \eta \Gamma} Q_{\alpha \beta} Q_{\beta \gamma} Q_{\gamma \alpha} - \frac{\xi\zeta}{4 \eta \Gamma}\left(Q_{\alpha \beta}^{2}\right)^{2}.
\end{align}

Note that this effective free energy is multiplied by $\xi \zeta$, implying that this product needs to be positive to favour the nematic state. The passive bulk free energy favours the isotropic state as $\gamma<\gamma_{NI}$ while the effective free energy favours the nematic state for extensile nematics of rod-like particles ($\zeta>0$ and $\xi>0$ respectively). By adding the passive and active free energies, $f_{NI}+f_A$, one can estimate the threshold activity where the nematic and isotropic states have the same effective free energy. The inset of Fig.~\ref{hf_zeta} (top) depicts the effective free energy at four different activities. At $\zeta=0$, there is only the passive term and the global minimum is at $S=0$ (isotropic). At $\zeta_{co}^*=0.00044$, both states have the same free energy. This ``coexistence'' activity can be calculated numerically by finding the minima of $f_{NI}+f_A$ and calculating the value of $\zeta$ for which they have the same value. At $\zeta=0.0001$, the global minimum is nematic and at $\zeta_i^*=0.022$, only a single nematic minimum remains. The activity, for which the isotropic state becomes unstable, may be calculated by setting the second derivative of $f_{NI}+f_A$ at $S=0$ to zero. Near $S=0$ only the quadratic term is relevant and the value of $\zeta_i^*$ is:
\begin{align}
    \zeta_i^* = \frac{A_0 \eta \Gamma}{\xi}(3-\gamma).
    \label{zeta2-eq}
\end{align}
The values of $\zeta_{co}^*$ and $\zeta_i^*$ as functions of $\gamma$ are plotted in Fig.~\ref{hf_zeta} (bottom) in green and orange, respectively. The threshold activity for interfacial propagation found in the simulations lies between these two limits, where the effective free energy has two locally stable minima, the isotropic state and the nematic which is lower. Thus, the interface propagates above $\zeta_{co}^*$, where the nematic state is favoured, and below $\zeta_i^*$ where the isotropic minimum loses stability. For $\zeta>\zeta_i^*$, the isotropic state is unstable and the system becomes nematic without a propagating interface. As one of the states is metastable, the interfacial propagation will also depend on the initial configuration as analyzed in the next section.

\begin{figure*}[htb]
\includegraphics[width=0.95\textwidth]{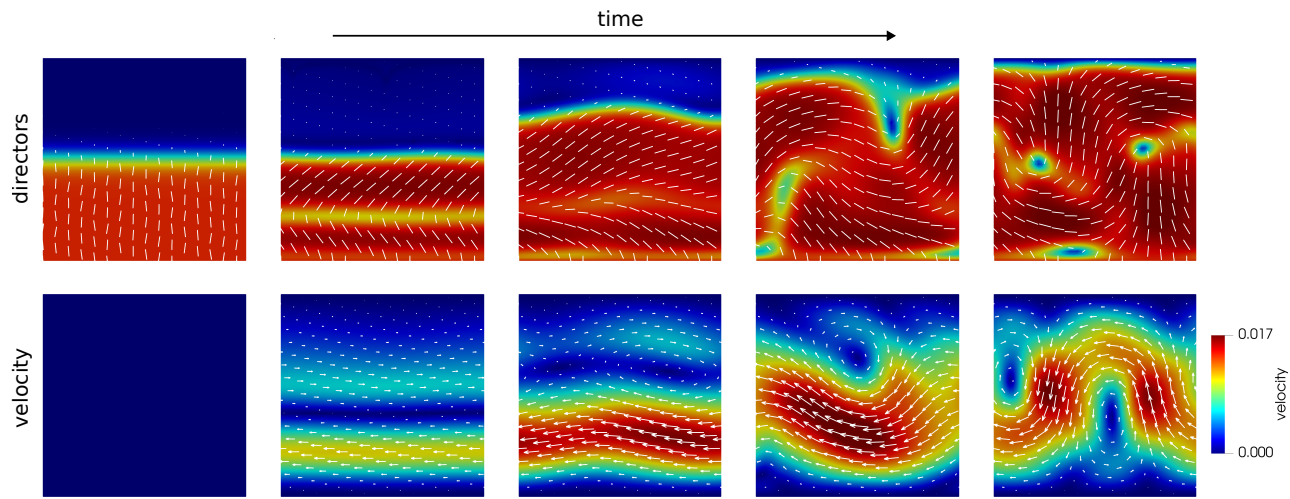}
\caption{Snapshots of the director and velocity fields at $\gamma=2.695$ and $\zeta=0.009$. The colors on the top represent the scalar order parameter with blue being isotropic and red nematic. The white lines represent the director field. On the bottom, the colours represent the magnitude of the velocity as indicated by the colour bar while the arrows indicate the direction of the velocity field.}
\label{2d-int-2}
\end{figure*}

\begin{figure*}[htb]
\includegraphics[width=0.95\textwidth]{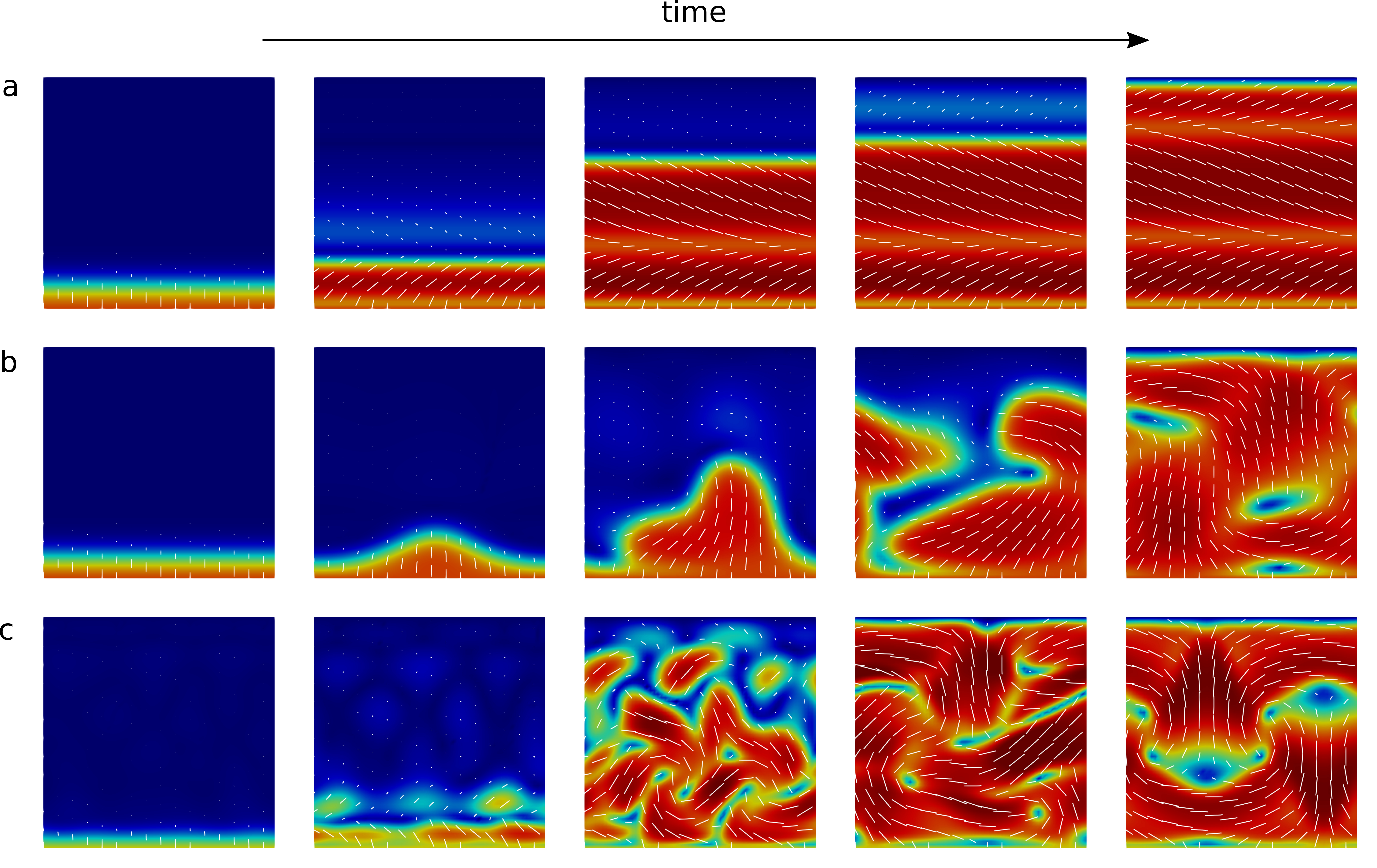}
\caption{Propagation of 2D interfaces at $\gamma=2.699$. a) 1D-like interface formed after a uniform isotropic initial condition and $\zeta=0.015$. b) Undulated interface after an initial condition with random fluctuation between $S=0$ and $0.05$ and at $\zeta=0.009$. c) Same as b, but with $\zeta=0.05$ (above $\zeta^*_i$). The colors represent the scalar order parameter with blue being isotropic and red nematic. The white lines represent the director field.}
\label{2d-int-3}
\end{figure*}

\subsubsection{Two-dimensional interface}

In the previous sections, we imposed translational invariance along the $x$-axis and, thus, the fields were allowed to vary only along the $z$-axis. As discussed before, this is adequate for the passive case. When activity is considered, for the same initial configuration used in 1D (isotropic fluid) the 2D simulations yield similar results. The 1D behaviour is observed even at relatively high activities.  Figure~\ref{2d-int} (left) illustrates the final configuration at $\zeta=0.009$ and $\gamma=2.695$. This 1D behaviour is a result of the simulation boundary and initial conditions (periodic boundaries and perfectly aligned directors at the surface), which are difficult to reproduce in experiments where the fluctuations are a rule and turbulent or pre-turbulent nematic states are more likely to occur. 
In the active case, the initial conditions are relevant and will determine the steady state. For instance, for the same parameters, when the system is initialized with a nematic-isotropic interface in the middle of the domain, with a homeotropic nematic with random director fluctuations of $10^\circ$, the interface propagates to the top as shown in Fig.~\ref{2d-int} (right). The fluctuations break the symmetry in the $x$-direction and promote the formation of defects and vortices. This dependence of the steady state on the initial conditions was not observed in the simulations of the passive system. 

One can rationalize this dependence on the initial conditions using the results of Fig.~\ref{hf_zeta}(bottom). At $\gamma=2.695$, 1D simulations with initial isotropic conditions yield a threshold activity for interfacial propagation, $\zeta^*=0.012$. Thus, as $\zeta=0.009 < \zeta^*$ (Fig.~\ref{2d-int} (left)), the interface does not propagate. However, this activity lies between $\zeta_{co}$ and $\zeta_i$, where the isotropic state is metastable and this is why a different initial condition results in interfacial propagation.

Figure~\ref{2d-int-2} illustrates the time evolution of the order parameter and the velocity field for the system considered in Fig.~\ref{2d-int} (right). The initial fluctuations of the director field nearly vanish and the nematic becomes uniformly aligned vertically. Then, a bending instability appears in the nematic accompanied by flow. After that, vortices are formed and the interface undulates. Eventually, the domain becomes entirely nematic and turbulent or pre-turbulent with defects created and destroyed repeatedly. Often the defects are formed at the interface as illustrated in Figure~\ref{2d-int-2} and previously reported in~\cite{C9SM00859D}. The final state resembles the dancing state observed in straight channels~\cite{Hardoin2019}, which is a set-up similar to that used here, although with different anchoring conditions (in the turbulent state, the anchoring at the surfaces has little influence in the defect dynamics).

Figure~\ref{2d-int-3}a illustrates the propagation of the 1D-like interface while
Figs.\ref{2d-int-3}b and c illustrate interfacial propagation in systems initialized with random fluctuations in the orientational order parameter $S$, ranging from $0$ to $0.05$ (isotropic state with fluctuations). For the lowest activity, $\zeta=0.009$, which is below $\zeta_{i}^*$, the interface undulates and propagates until the domain becomes active turbulent. For $\zeta=0.05$, which is above $\zeta_i^*$, the domain becomes nematic from the bulk, i.e., before a film is clearly formed and the interface propagates, as the isotropic phase is unstable. At early times there is still an ordered region close to the surface as a result of the strong anchoring conditions but this region is inhomogeneous and an interface between the ordered and disordered regions is not clearly visible.   

\section{Conclusion}
\label{conclusion-sec}

We investigated how a nematic-isotropic interface near a flat ordering surface is affected by three factors: the anchoring strength, the ordering field and the activity. We have used an analytical expression for the orientational order parameter profile of the passive system to validate the results of the numerical simulations and used the latter to study the steady states and the dynamics of the interfaces in active systems. The simulations were based on the Beris-Edwards equation for nematohydrodynamics. 

In the passive case, we found that the anchoring does not affect the interfacial height except for very weak anchorings. Above a certain anchoring, the behaviour is the same as that for infinite anchoring, away from the nematic-isotropic transition. When the ordering field approaches the nematic-isotropic transition, the interfacial height diverges logarithmically for infinite systems as an orientational wetting transition occurs at these anchorings. 

For active interfaces, we observed two regimes. Below a certain threshold activity, the interfaces behave as the passive ones and, above it, they propagate until the entire domain becomes active nematic. We found that this threshold activity decreases as the ordering field approaches the transition value. Furthermore, the final steady state for active interfaces is very sensitive to the initial configuration, by contrast to the passive case. We interpret this dependence on the initial conditions in terms of the effective free energy generated by the activity, which favours the nematic phase for extensible systems of rod-like particles. When the system is initially isotropic, the interface is essentially one-dimensional. If a perturbation that breaks the translational symmetry is introduced, the problem becomes two-dimensional with the formation of defects, vortices and undulated interfaces, eventually leading to states in the pre-turbulent or turbulent active nematic regimes. 

\section*{Acknowledgements}

We acknowledge financial support from the Portuguese Foundation for Science and Technology (FCT) under the contracts: PTDC/FISMAC/5689/2020, EXPL/FIS-MAC/0406/2021, UIDB/00618/2020, UIDP/00618/2020 and DL 57/2016/CP1479/CT0057. This work was produced with the support of UC-LCA and it was funded by FCT I.P. under the project Advanced Computing Project 2023.09574.CPCA.A1, platform Navigator.

\bibliography{biblio}

\end{document}